\documentclass[aps,prd,reprint,superscriptaddress]{revtex4-2}
\usepackage{graphicx,setspace}
\usepackage{dcolumn}
\usepackage{bm,url}
\usepackage[mathlines]{lineno}
\usepackage[linktocpage,colorlinks=true,citecolor=teal,linkcolor=olive,anchorcolor=green,urlcolor=teal]{hyperref}
\usepackage{mathrsfs}
\usepackage{xcolor}
\usepackage{float}
\usepackage[normalem]{ulem}
\usepackage[utf8]{inputenc}
\usepackage{tensor}
\usepackage{physics}
\usepackage{blindtext}
\usepackage{slashed}
\usepackage{tikz}
\usepackage{amsthm,amsfonts,epsfig,natbib,amssymb}
\usepackage{tensor}
\usepackage{mathtools}
\usepackage[scaled]{beramono}
\usepackage[T1]{fontenc}
\usetikzlibrary{decorations.pathmorphing}

\def\be{\begin{equation}}
\def\ee{\end{equation}}

\begin{document}
\title{Collisional Penrose process of extended test particles near an extremal Kerr black hole}

\author{Aofei Sang}
\email[]{aofeisang@mail.bnu.edu.cn}
\affiliation{Department of Physics, Southern University of Science and Technology, Shenzhen 518055, China}
\author{Jie Jiang}
\email[]{jiejiang@mail.bnu.edu.cn}
\affiliation{College of Education for the Future, Beijing Normal University, Zhuhai 519087, China}
\author{Ming Zhang}
\email[]{mingzhang@jxnu.edu.cn, corresponding author}
\affiliation{Department of Physics, Jiangxi Normal University, Nanchang 330022,China}

\begin{abstract}
We investigate the collisional Penrose process of extended test particles near an extremal Kerr black holes using the pole-dipole-quadrupole approximation. We analyze the motion of the test particles and examine the dynamics and maximum efficiency of energy extraction in this process. Our results demonstrate that the maximum extracted energy in the collisional Penrose process is influenced by the spin $s$ and quadrupolar parameter $C_{ES^2}$ of the test particles. Specifically, we observe that, at a fixed collisional position, the energy extraction efficiency decreases as the spin increases for either the pole-dipole or the pole-dipole-quadrupole approximation case. Furthermore, for a fixed spin, the energy extraction efficiency is higher in the pole-dipole-quadrupole approximation compared to the pole-dipole approximation. These findings provide insight into the role of the internal structures of the test particles in the collisional Penrose process.
\end{abstract}

\maketitle

\section{Introduction}

The Penrose process, discovered by Roger Penrose in 1969, offers a captivating approach to extracting energy from rotating black holes  \cite{Penrose:1971uk}. In the vicinity of {the rotating} black holes, within the ergoregion, particles exhibit intriguing properties, including the possibility of negative energies as observed from distant vantage points. Penrose proposed that an object with initial energy can fragment into two distinct parts: one part escapes to infinity with higher energy, while the other is absorbed by the black hole. This unique process leads to a net gain in energy at the expense of the rotational energy of the black hole. However, the efficiency of the Penrose process is inherently limited, with a maximum efficiency of approximately {120.7\%} achieved when the object disintegrates into two particles  \cite{Bardeen:1972fi,Wald:1974kya}. Despite this limitation, the Penrose process provides invaluable insights into the intricate behavior of energy in the presence of rotating black holes, contributing to our understanding of astrophysical phenomena  \cite{Fields:2014pia,Gondolo:1999ef}.

While the original Penrose process yielded only modest energy gains, subsequent research has turned its attention to a variant known as the collisional Penrose process  \cite{piran1975high,Harada:2014vka,Berti:2014lva}. Particle collisions within the ergoregion of rotating black holes have been extensively investigated, particularly in the context of {dark matter} searches \cite{Berti:2014lva,Banados:2009pr}. Recent studies have suggested that the efficiency of this process, quantified by the ratio {$\eta$} between the energy of post-collision particles detected at infinity and the energy of the colliding particles, should exhibit modest values $(\eta \lesssim 1.5)$ \cite{Harada:2012ap,Bejger:2012yb}. However, ongoing research has unveiled the potential for significantly higher maximum efficiencies, opening up new possibilities for extracting more substantial amounts of energy and unraveling intriguing astrophysical phenomena \cite{Berti:2014lva}.

Traditionally, research has primarily focused on point-like or pole-dipole particle models \cite{Hojman:1976kn,Zhang:2020cpu}. However, as research progresses, considering the more complex internal structure  of particles has become increasingly important \cite{Steinhoff:2012rw,Bini:2013rrx,Zhang:2022tlk,Timogiannis:2023pop,Hojman:1976kn,Zhang:2020cpu,Papapetrou:1951pa,Steinhoff:2009tk}.
One important model in this regard is the quadrupole body, which can describe the finite size and internal structure of particles.
The study of quadrupole bodies has attracted significant interest, particularly in the context of phenomena related to {gravitational wave physics and black hole dynamics} \cite{Papapetrou:1951pa,Steinhoff:2009tk}. This is because the quadrupole body model provides a more realistic description of objects compared to traditional point-like {or pole-dipole} models, allowing for a more accurate representation of particle spin, deformations, and gravitational interactions. {Investigating the phenomena involving the quadrupole bodies contributes to our understanding of the behavior of objects near black holes and their role in gravitational wave detection.}

In this paper, we would like to consider the collisional Penrose process involving test particles with internal structure, specifically focusing on the pole-dipole- quadrupole approximation.  In Sec. \ref{sec2}, we present the equation of motion for the particle and discuss the conserved quantities during the motion. In Sec. \ref{sec3}, we introduce the collisional Penrose process and discuss the conditions under which this process can occur. In Sec. \ref{sec4}, we present numerical results and provide a detailed analysis {on the maximal extracted energy as well as the maximum efficiency}. Finally, in Sec. \ref{sec5}, we summarize our findings and draw conclusions.

\section{Equations of motions and conserved quantities of an extended test particle }\label{sec2}

In a framework of curved spacetime, the trajectory of a particle is influenced by the Lorentz-like force arising from its intrinsic spin, deviating it from a purely geodesic motion. The description of an extended test particle traversing through such curved spacetime necessitates the inclusion of the particle's multipole momentum up to the quadrupolar level. This is achieved through the utilization of the Mathisson-Papapetrou-Dixon (MPD) equations in their pole-dipole-quadrupole approximation form  \cite{Steinhoff:2012rw,Steinhoff:2009tk}:
\begin{equation}\begin{aligned}\label{mpdequ}
&\frac{D p_a}{D\tau}=-\frac{1}{2}R_{abcd}u^{b}S^{cd}-\frac{1}{6}\nabla_{a} R_{bcde}J^{bcde}\,,\\
&\frac{D S^{ab}}{D \tau}=2 p^{[a}u^{b]}-\frac{4}{3}R_{cde}{}^{[a}J^{b]ecd}.
\end{aligned}\end{equation}
Here, the particle's four velocity, four-momentum and spin tensor are denoted by $u^a$, $p^a$ and $S^{ab}$, respectively. The four-velocity $u^a\equiv {\rm{d}} x^{\mu} / {\rm{d}} \tau$ can be normalized to $u^a u_a=-1$ via a suitable choice of trajectory parametrization  \cite{Obukhov:2010kn}. The total derivative along the particle's world line is denoted by ${D}/{D}\tau=u^{a} \nabla_{a}$.
The spin tensor $S^{ab}$ is antisymmetric.
And $J^{abcd}$ stands for the quadrupole tensor, which adheres to \cite{Steinhoff:2012rw,Steinhoff:2009tk}
\begin{equation}
J^{a b c d}=J^{[a b][c d]}=J^{c d a b},
\end{equation}
and
\begin{equation}
J^{[a b c] d}=0 \Leftrightarrow J^{a b c d}+J^{b c a d}+J^{c a b d}=0\,,
\end{equation}
i.e. the quadrupole tensor shares the same algebraic symmetries with the Riemann tensor $R_{abcd}$. Then, we need to introduce the spin supplementary condition (SSC)\cite{tulczyjew1959motion,dixon1964covariant,Dixon:1970zza} because the particle's dynamics cannot be entirely predicted by the MPD equations and the four-velocity's normalization condition.  In this paper, we employ the covariant Tulczyjew-Dixon SSC:
\begin{equation}\label{sscc}
S^{ab} p_{a}=0.
\end{equation}
This condition, frequently applied to extended test particles, introduces three additional equations that facilitate the fixation of the reference worldline.

With this condition, we can define the relationship between $S_{ab}$ and a spacelike spin four-vector $S^a$ by
\begin{equation}\label{Sab}
S^{ab}=-\frac{1}{m}\epsilon^{abcd} S_{c} p_{d},
\end{equation}
where $\epsilon_{abcd}$ is the Levi-Civita tensor. Then, we would like to define the spin length $S$ by
\be\begin{aligned}\label{spinlength}
    S^{ab}S_{ab}=2S^a S_a=2 S^2.
\end{aligned}\ee
For quadrupolar tensor, in this paper, we would like to employ a specialized model, designed explicitly to analyze a rotating black hole. This model is denoted by  \cite{Steinhoff:2012rw,Hinderer:2013uwa},
\begin{equation}\label{ssqm}
J^{abcd}=-\frac{3}{m^2} p^{[a} Q^{b][c} p^{d]},
\end{equation}
where $m^2$ is defined as $m^2\equiv -p^a p_a$. Here, the tensor $Q_{ab}$ is expressed as
\begin{equation}\label{qab_10}
Q_{ab}=c_{ES^2} S_{ac} S_{b}{}^{c}\,,
\end{equation}
where $c_{ES^2}$ is a coupling constant which is related to the quadrupolar deformations {induced by spin\cite{Steinhoff:2012rw,Porto:2008jj}}.

It is convenient to find conserved quantities when we study the motion of the extended particle. First, upon employing the spin-induced quadrupole momentum tensor, expressed in Eq. (\ref{ssqm}), and the Tulczyjew-Dixon SCC (\ref{sscc}), we can establish that  \cite{Steinhoff:2012rw}
\begin{equation}\label{vanis}
\mathrm{D}S/\mathrm{D}\tau=0.
\end{equation}
This equation implies that the magnitude of the spin $S$ is conserved under these particular conditions.

Then, by referring to Eq. (\ref{mpdequ}), we obtain
\begin{equation}
p^a=-\frac{{\rm{D}} S^{a b}}{{\rm{D}} \tau}u_b+m_0 u^a-\frac{4}{3} R_{cde}{}^{[a} J^{b] ecd}u_b,
\end{equation}
where the particle's rest mass is denoted by $m_0\equiv -p^a u_a$. This leads to the conclusion that
\begin{equation}\label{eq13}
\begin{aligned}
m^2&=\frac{{\rm{D}} S^{ab}}{{\rm{D}} \tau}u_b p_a+m_0^2+\frac{4}{3} R_{cde}{}^{[a} J^{b] ecd}u_b p_a,
\end{aligned}
\end{equation}
and
\begin{equation}\label{eqmpsr}
\begin{aligned}
\frac{{\rm{D}} m}{{\rm{D}} \tau}=\frac{m}{6 m_0} \frac{{\rm{D}} R_{abcd}}{{\rm{D}} \tau} J^{abcd}+O(\epsilon^3).
\end{aligned}
\end{equation}
This implies that the mass parameter, $m$, does not remain constant when considering the pole-dipole-quadrupole approximation of the test particle.
To control the variables, we need a mass which is conserved in the motion under the pole-dipole-quadrupole approximation.
Building upon previous research \cite{Steinhoff:2012rw}, a perturbatively conserved mass parameter can be formulated as
\begin{equation}\begin{aligned}\label{conservedmass}
\mu=m-\frac{1}{2 m^2}E_{bc} Q^{bc}+O(\epsilon^3)\,,
\end{aligned}\end{equation}
where $E_{bd}=R_{abcd} p^a p^c$. By applying the relation $\frac{D S^{ab}p_{b}}{D\tau}=0$ and substituting $m$ with $\mu$, a connection can be established between the 4-momentum $p^a$ and the 4-velocity $u^a$  \footnote{{It is worth to note that the relation between $u^a$ and $p^a$ given in Ref.\cite{Hinderer:2013uwa} is obviously wrong in the dimension because they set the dimensionful parameter $c_{ES^2}$ as $1$ by no reason. Here, we recalculate this relation and give a right one.}}:
\begin{equation}\begin{aligned}\label{relationup}
u^a=&\frac{p^a}{\mu}\left(1+\frac{1}{2 \mu^3}Q^{bc}E_{bc} \right)-\frac{1}{2\mu^3}R_{bec}{}^{d}p^e S^{ba}S^c{}_d\\
&+\frac{1}{\mu^2}R_{cde}{}^a Q^{ec}p^d+\frac{1}{\mu^4}Q^{bc}R_{cde}{}^a p_b p^e p^d\,.
\end{aligned}\end{equation}

Another conserved quantity is given by
\begin{equation}\label{Exi}
    E_{\xi} = p_a \xi^a + \frac{1}{2} S^{ab} \nabla_a \xi_b\, .
    \end{equation}
According to  \cite{Steinhoff:2009tk}, in the pole-dipole case where $J^{abcd} = 0$, this quantity
is preserved if $\xi^a$ is a Killing-vector, with $\nabla_{(b} \xi_{a)} = 0$. Moreover, from  \cite{ehlers1977dynamics}, this holds true as a preserved quantity at all higher multipole orders including quadrupolar case.

In this paper, we would like to focus on {the} Kerr black hole. The metric in Boyer-Lindquist coordinate is
\begin{equation}\label{metric}
\begin{aligned}
\rm{d}s^2=&-\frac{\Delta}{\Sigma}\left({{\rm{d}} t}-{a \sin ^{2} \theta} {\rm{d}} \phi\right)^{2}+\frac{\Sigma}{\Delta} {\rm{d}} r^{2}+{\Sigma} {\rm{d}} \theta^{2}\\
&+\frac{\sin^{2} \theta}{\Sigma}\left({a {\rm{d}} t}-{(r^{2}+a^{2})} {\rm{d}} \phi\right)^{2},
\end{aligned}
\end{equation}
{where}
\begin{align}
\Sigma &=r^2+a^2 \cos ^2\theta,\nonumber~\\
\Delta &=\left(r^{2}+a^2\right)-2 Mr.\nonumber\
\end{align}
The ergosphere is given by
\begin{equation}\begin{aligned}
&r_+< r< r_+^s,\\
&r_+=M+\sqrt{M^2-a^2},\\
&r_+^s=M+\sqrt{M^2-a^2 \cos^2\theta}.
\end{aligned}\end{equation}
Here, $r_+$ denotes the outer event horizon, and $r_+^s$ signifies the outer ergosphere. $M$ and $a$ are the black hole's mass and per-unit-mass angular momentum, respectively. {For brevity}, we will assume $M=1$.

Taking into account the symmetry, it's hypothesized that collisions occurring on the equatorial plane have the potential to generate particles of maximal energy, as proposed in Ref. \cite{Bejger:2012yb}. Consequently, our subsequent analyses will concentrate on movements within the equatorial plane, that is,
\begin{equation}
\theta=\frac{\pi}{2},\quad p^\theta=0.
\end{equation}
Moreover, we'll delve into scenarios where the spin is aligned in relation to the rotating background source, characterized by
\begin{equation}
S^{a\theta}=0.
\end{equation}
Considering the definition \eqref{Sab} in conjunction with the above assumptions, it's apparent that $S^a$ has only one component that is non-zero:
\begin{equation}
S^a=S^{\theta}\delta^a_{\theta}.
\end{equation}
By applying the relation $2S^a S_a=-S_{ab}S^{ab}$ along with the definition of the {spin length from equation \eqref{spinlength}}, we arrive at:
\begin{equation}
-S^{\theta}=\frac{S}{\sqrt{g_{\theta\theta}}}.
\end{equation}
In the following discuss, we denote the spin direction by the sign of $S$. A detailed contemplation on the spin orientation is provided in  \cite{Steinhoff:2012rw}. This subsequently allows us to express the components of the spin tensor in terms of the spin length {as}
\begin{equation}
\begin{aligned}
&S^{rt}=-\frac{S p_{\phi}}{m r}=-\frac{\mu s p_{\phi}}{m r}\,,\\
&S^{\phi t}=\frac{S p_r}{m r}=\frac{\mu s p_r}{m r}\,,\\
&S^{\phi r}=-\frac{S p_t}{m r}=-\frac{\mu s p_t}{m r}\,,
\end{aligned}
\end{equation}
where we have {constrained} our position within the equatorial plane and rescale $S$ by $s=S/\mu$.

In {the} Kerr spacetime, there are two Killing vectors: {$\xi^a_{t}=(\partial/\partial t)^a$ and $\xi^a_{\phi}=(\partial/\partial \phi)^a$}. The corresponding conserved energy $E$ and conserved angular momentum $J$ {of the particle} can be found by replacing the Killing vectors into \eqref{Exi}:
\be\begin{aligned}
    E= p_a \xi_t^a + \frac{1}{2} S^{ab} \nabla_a \xi_t{}_b,\\
    J= p_a \xi_{\phi}^a + \frac{1}{2} S^{ab} \nabla_a \xi_{\phi}{}_b.
\end{aligned}\ee
Here, we would like to rescale $E$ and $J$ by $e=E/\mu$ and $j=J/\mu$.
Then, we can express $p^t$ and $p^\phi$ in terms of the constants of motion. {As a result,} corresponding with $-m^2=p^a p_a$ and \eqref{conservedmass}, we can find
\be\begin{aligned}
    m=& \mu- s^2\frac{c_{ES^2} \mu^2 M \left(3 j^2-6 e j a+3 e^2 a^2+r^2\right)}{2 r^5},\\
p^t=&\frac{\mu}{r\Delta}\left(-2 j M a +2 e M a^2+e a^2 r+e r^3\right)\\
&+\frac{\mu s}{r^3 \Delta}\left[j M (a^2+r^2)-e M (a^3+3 a r^2)\right]\\
&+\frac{\mu s^2}{r^4\Delta}\left[-2 j M^2 a+e\left(2 M^2 a^2+M r(a^2+r^2)\right)\right],\\
p^\phi=&\frac{\mu}{r \Delta}\left(-2 j M+2 e M a+j r\right)\\
&+\frac{\mu s}{r^3 \Delta}\left(j M a-e M a^2-2 e M r^2+e r^3\right)\\
&+\frac{\mu s^2}{\Delta}\left(-2 j M^2+2 e M^2 a+j M r\right)
\end{aligned}\ee
under pole-dipole-quadrupole approximation within the equatorial plane for the particle.

According to $p^a p_a=-m^2$, we can find
\be\begin{aligned}
    (p^r){}^2=\frac{1}{g_{rr}}\left[-m^2-g_{tt}(p^t)^2-g_{\phi \phi}(p^\phi)^2-2 g_{t\phi}p^{\phi}p^t\right].
\end{aligned}\ee
Letting $p^r=\sigma\sqrt{O}$ with $\sigma=\pm 1$. {{ When $\sigma=1$, the particle is outgoing and the particle is ingoing when $\sigma=-1$.}} Considering the quadrupolar approximation, we can find the explicit expression of $O$. It is easy to notice that $O$ must be non-negative. Then, using the relation between $u^a$ and $p^a$ \eqref{relationup}, we can get the 4-velocity {$u^a$ for the particle}.

\section{collisional penrose process of the particle with the pole-dipole-quadrupole approximation}\label{sec3}

In this paper, we consider the collisional penrose process of around the Kerr black hole. {According to Ref. \cite{Berti:2014lva}, the super Penrose process will occur only when the black hole is extremal. Therefore, we would focus on extremal black hole with $a=1$ in this paper.} We assume that there are 4 particles. Firstly, there is a particle 1, which is falling into the black hole and has a 4-momentum denoted by $p^{\mu}_{1}$, and particle 2, which is away from the black hole with a 4-momentum of $p^{\mu}_{2}$.       These particles collide within the ergosphere of the black hole and result in the creation of particle 3, which escapes to infinity carrying a 4-momentum of $p^{\mu}_{3}$, and particle 4, which descends into the black hole, taking along a 4-momentum of $p^{\mu}_{4}$.

According to the above settings, we can find some constraints on the motion of these particles. We assume that the collisional point is at $r=r_c$. Firstly, it's important to note that particle 3 is bound to make its way to infinity, which suggests there is no turning point for particle 3 in the region where $r>r_c$, as demonstrated in Refs. \cite{Zhang:2018gpn,Berti:2014lva,Schnittman:2014zsa}. Specifically, we need
\be\begin{aligned}
    O\geq 0 \,,r\geq r_c
\end{aligned}\ee
for the particle 3 and similarly
\be\begin{aligned}
    O\geq 0 \,,r_c\geq r\geq r_+
\end{aligned}\ee
for the particle 4.
Then, we require that
\be\begin{aligned}\label{con2}
u^t > 0
\end{aligned}\ee
for particle $3$ and particle $4$ to ensure the causality and the local measured energy positive \cite{Berti:2014lva}.

The 4-momentum at the collision point is locally conserved, as pointed out in Refs. \cite{Zhang:2018gpn,Maeda:2018hfi}. This conservation law can be explicitly written in terms of energy $e$, angular momentum $j$, and radial momentum $p^r$ as
\begin{equation}
\begin{aligned}
e_3+e_4&=e_1+e_2,\\
j_3+j_4&=j_1+j_2,\\
\sigma_3 |p^r_3|+\sigma_4 |p^r_4|&=\sigma_1 |p^r_1|+\sigma_2 |p^r_2|,
\end{aligned}
\end{equation}
where $\sigma_1=-\sigma_2=-\sigma_3=\sigma_4=1$ represents the direction of the radial momentum.

In this scenario, we are particularly interested in a direct collision occurring in the ergosphere between two test particles. To simplify our analysis, we assume that the particles share the same mass, energy, angular momentum, and spin, that is, $\mu_1=\mu_2=\mu$, $e_1= e_2= e=1$, and $j_1= j_2= j$, $s_1 = s_2=s$. Further, for computational feasibility, we also assume $\mu_3=\mu_4=\mu$ and $s_3=s_4=s$. { It is worth to note that $s$ must satisfy $|s|\ll M$, i.e. $|s|\ll 1$, because the energy density of a body must be positive \cite{Guo:2016vbt}. In this paper, we would like to consider $s$ whose value ranges from $-0.1$ to $0.1$.}
{Besides, we also assume each particle share the same quadrupolar parameter $c_{ES^2}$. Then, once given the explicit numerical values of $s$, $j$, and
\begin{equation}
    C_{ES^2}:=\mu c_{ES^2}\,,
\end{equation}
we can find the maximal value of $e_3$ after considering the above constraint. And $C_{ES^2}$ reaches $1$ for a rotating black hole \cite{Steinhoff:2012rw}. 
Then, the maximum energy extract efficiency will be $\eta=e_{3max}/2$.} {Without loss of generality, we will only concentrate on $e_{3max}$ in the following discussion.} Moreover, in this paper, we will consider the case where $C_{ES^2}=0$ and $C_{ES^2}=1$, in which $C_{ES^2}=0$ represents for the quadrupolar effect being vanished and $C_{ES^2}=1$ for the quadrupolar effect being considered. 

\begin{figure}
    \centering
    \includegraphics[width=0.48\textwidth]{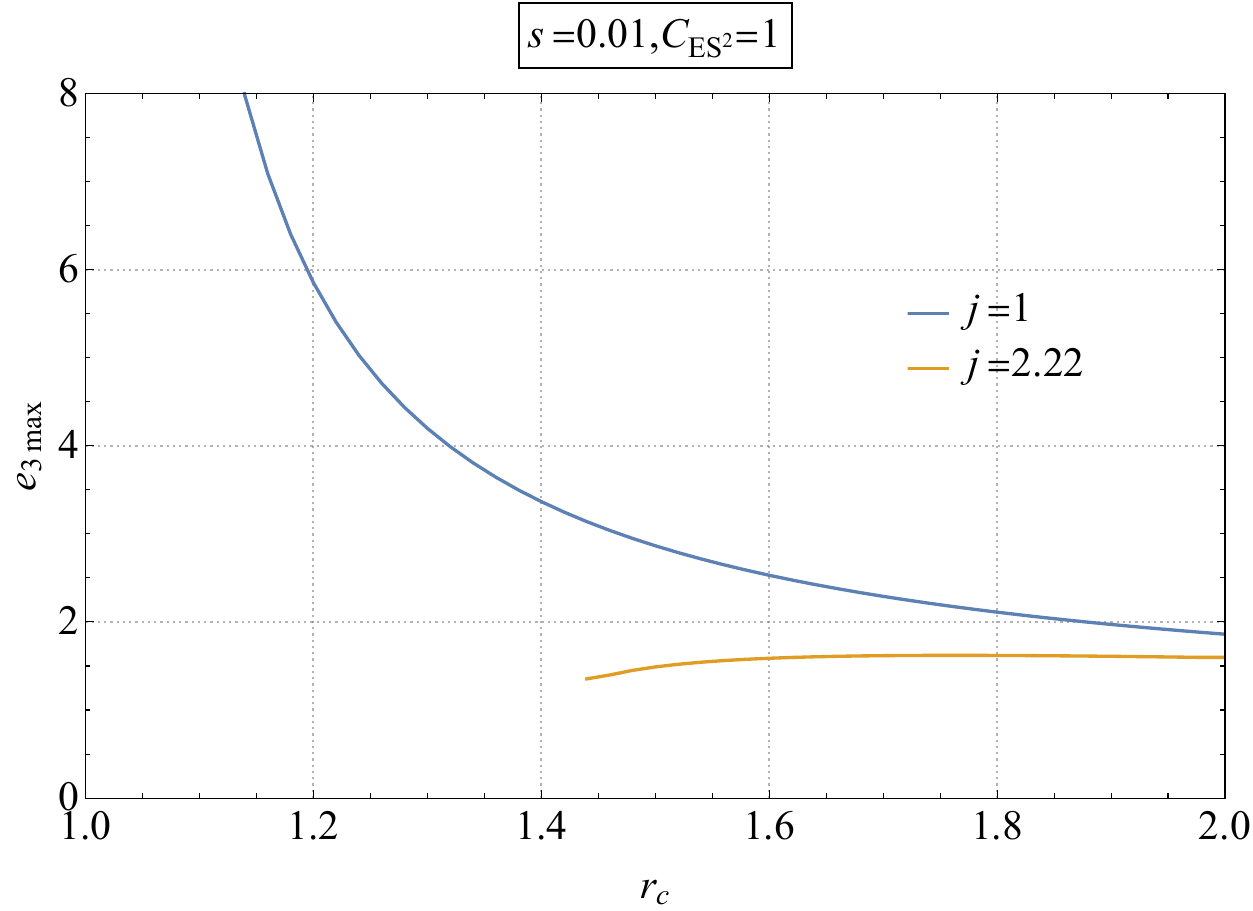}
    \caption{Maximum energy extracted by the collisional Penrose process as a function of collisional points $r_c$ for different $j$.}
    \label{fig1}
\end{figure}

\begin{figure*}
    \centering
    \includegraphics[width=0.48\textwidth]{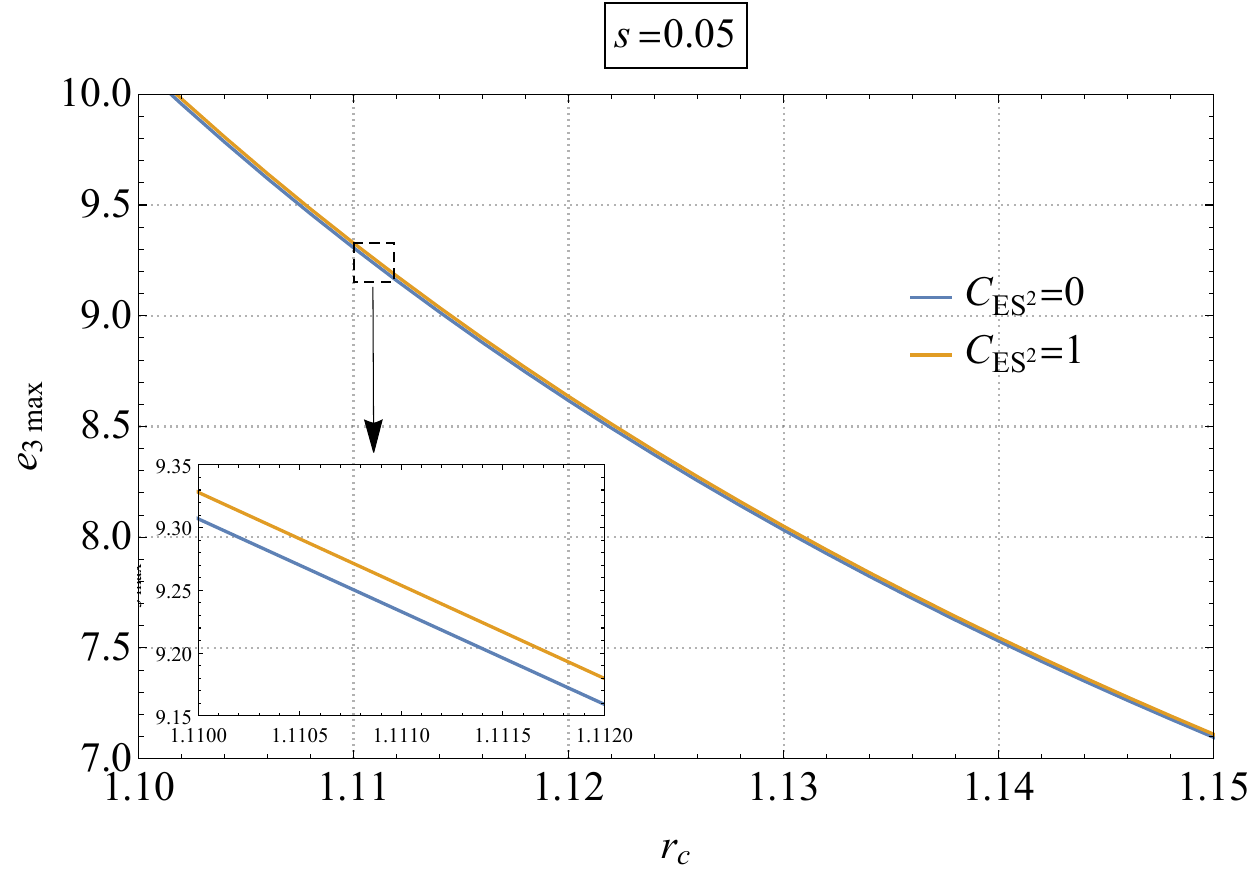}
 \includegraphics[width=0.48\textwidth]{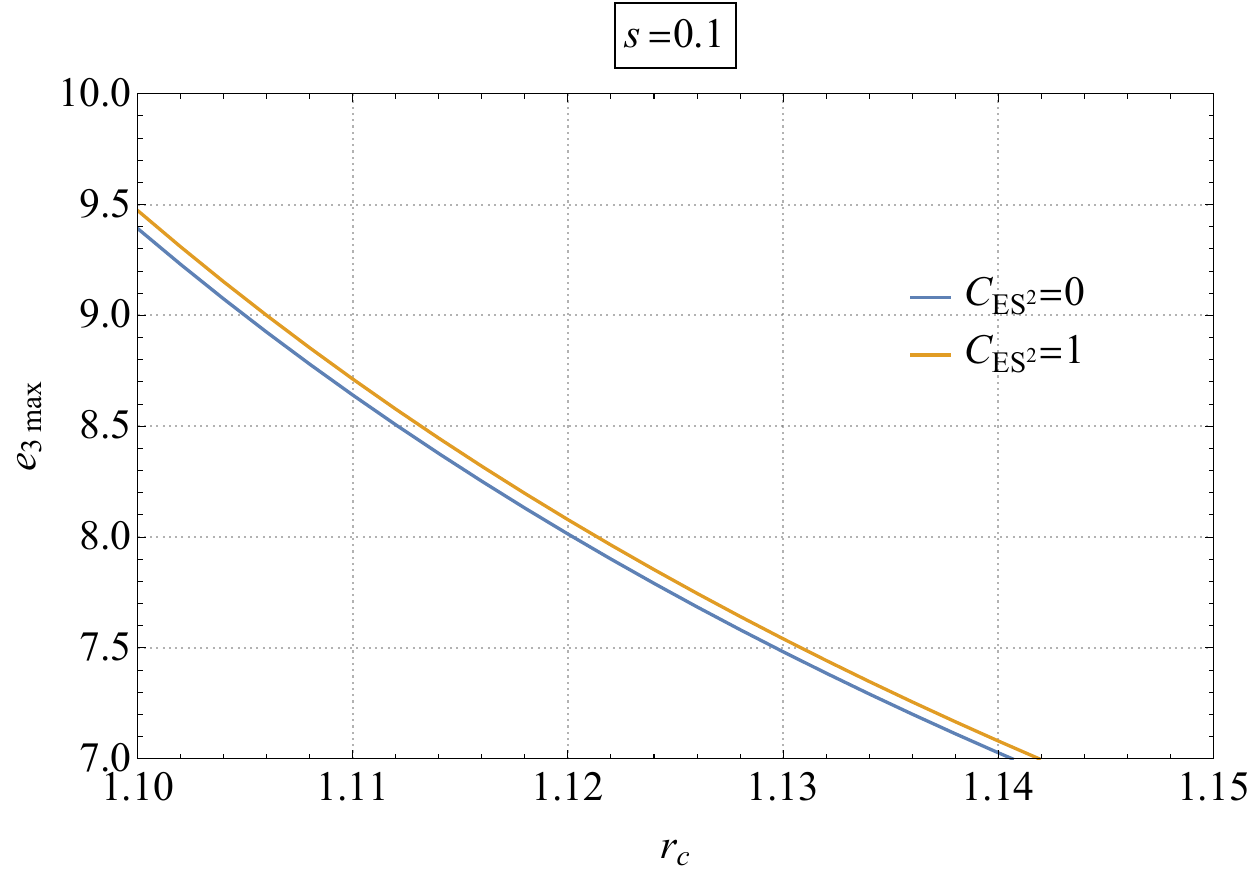}
 \includegraphics[width=0.48\textwidth]{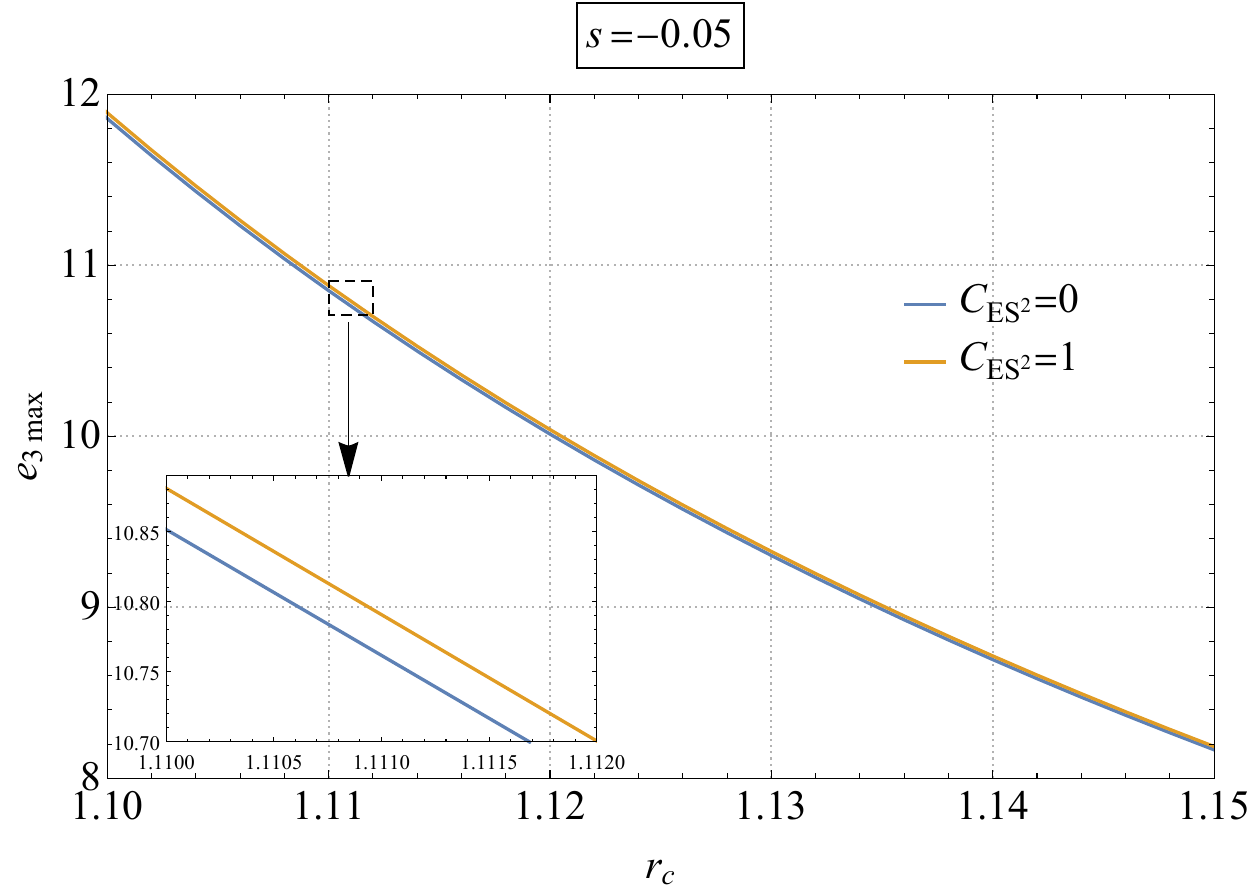}
 \includegraphics[width=0.48\textwidth]{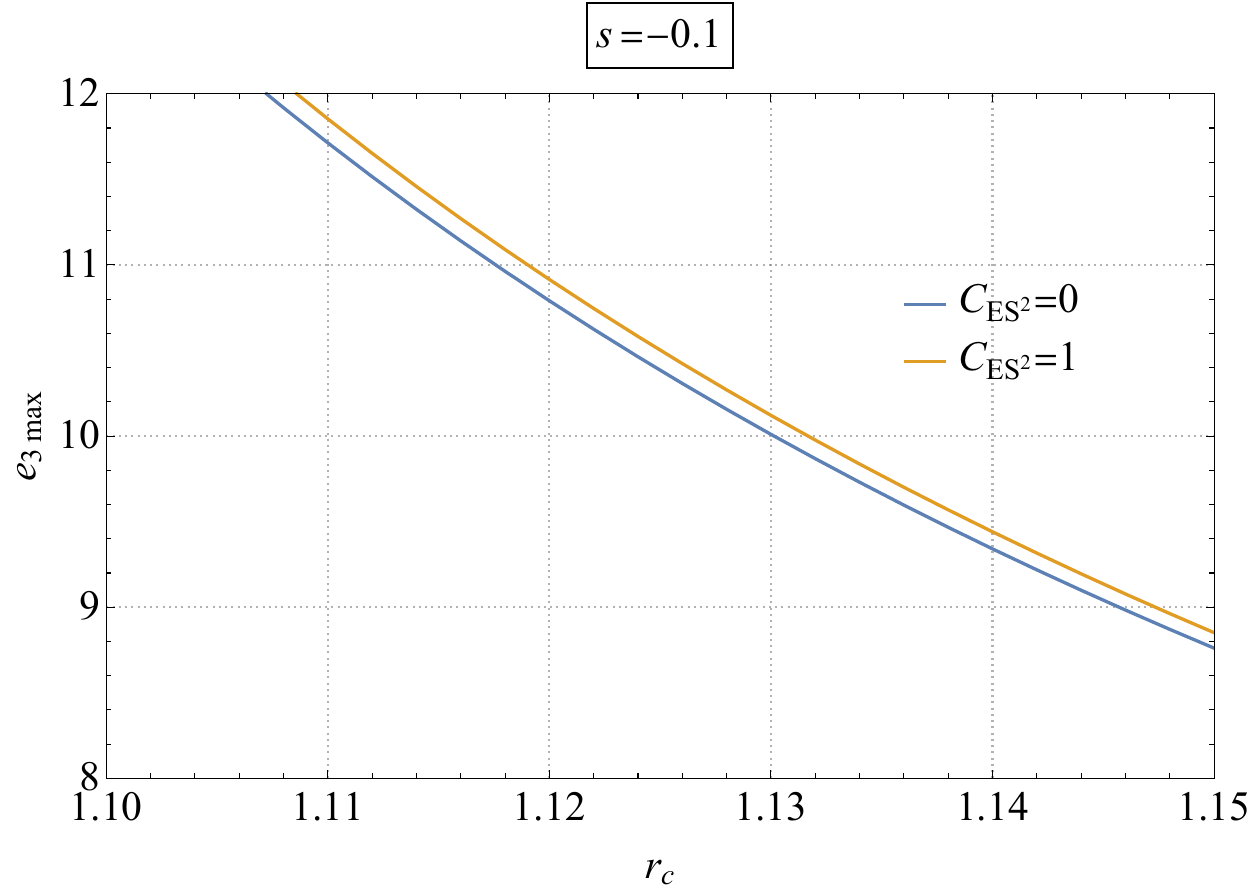}
    \caption{Maximum energy extracted by the collisional Penrose process as a function of the collision point for different spin parameters. Each panel shows both $C_{ES^2}=0$ and $C_{ES^2}=1$ cases.}
    \label{fig2}
\end{figure*}

\begin{figure*}
    \centering
    \includegraphics[width=0.48\textwidth]{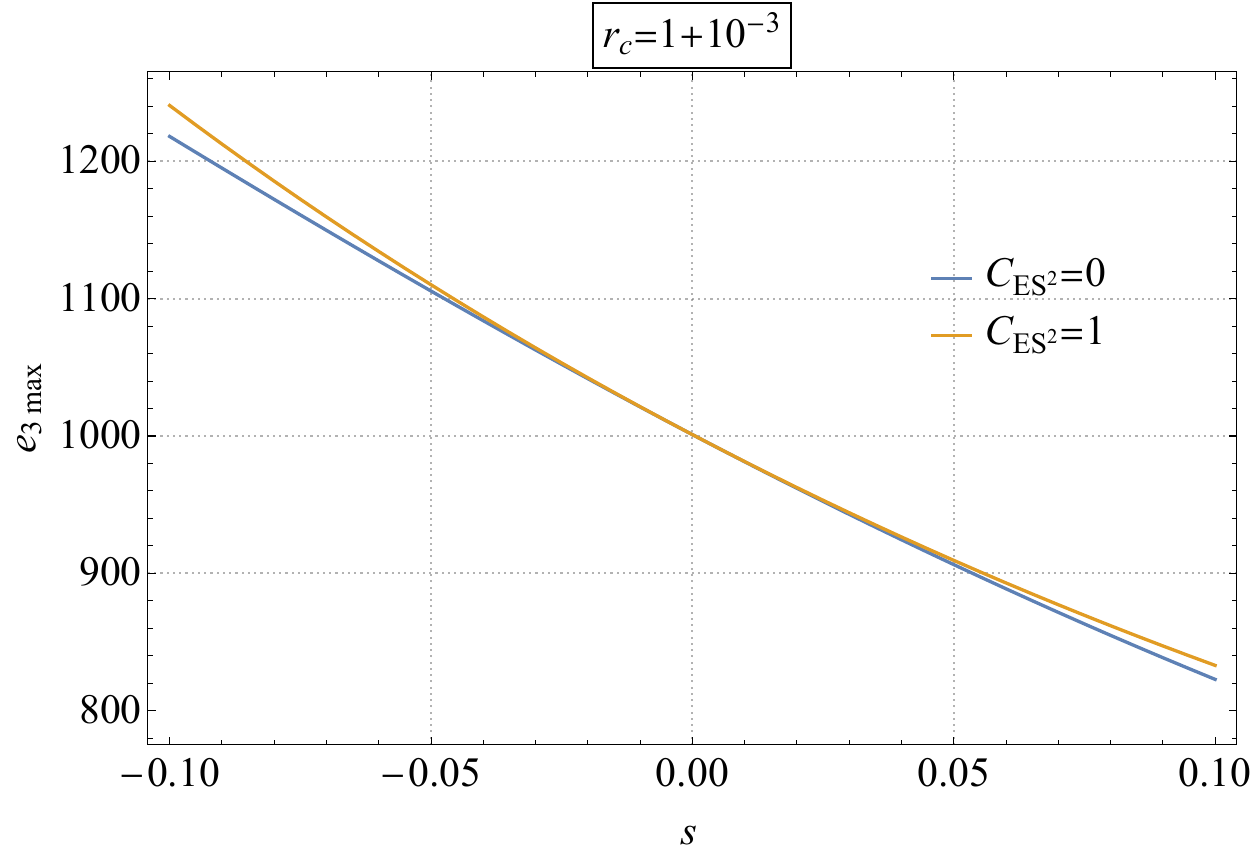}
 \includegraphics[width=0.48\textwidth]{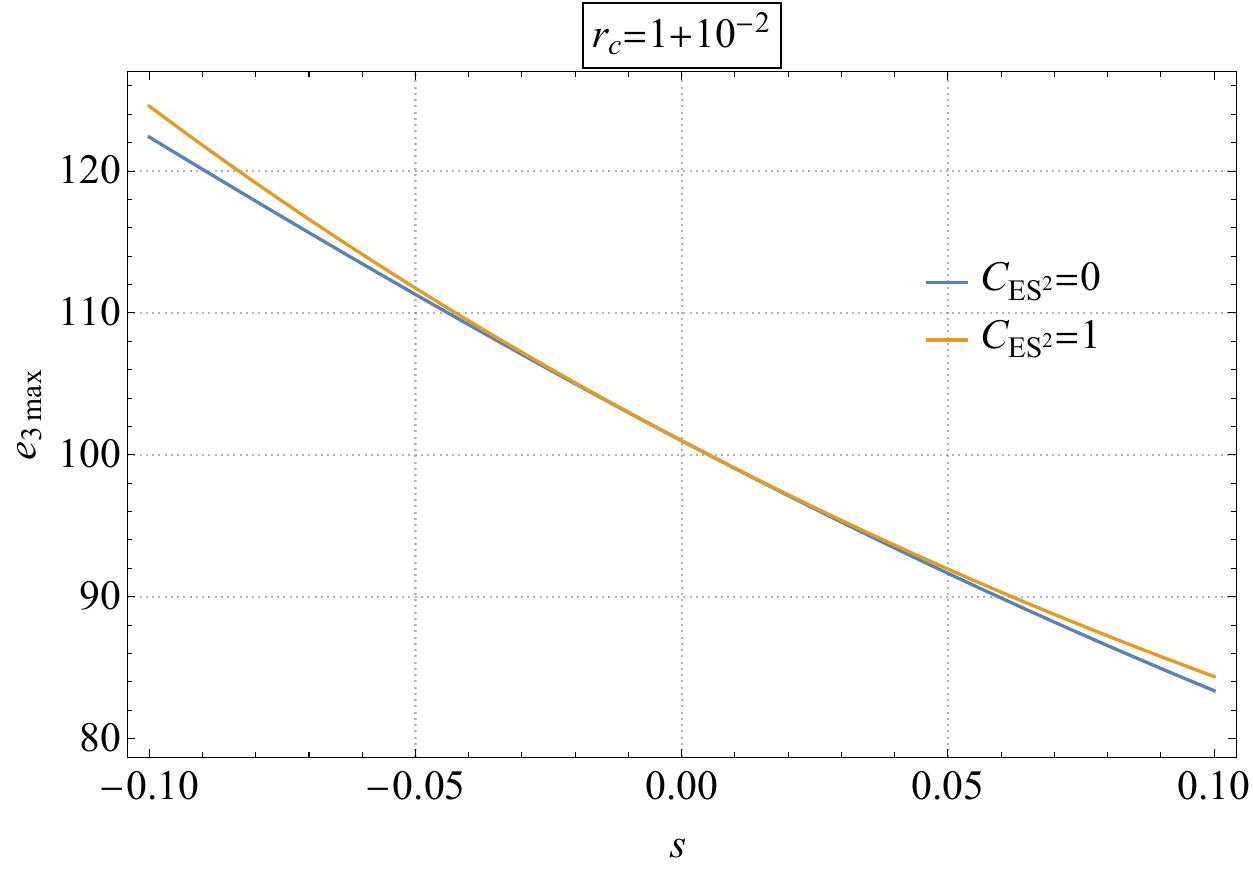}
 \includegraphics[width=0.48\textwidth]{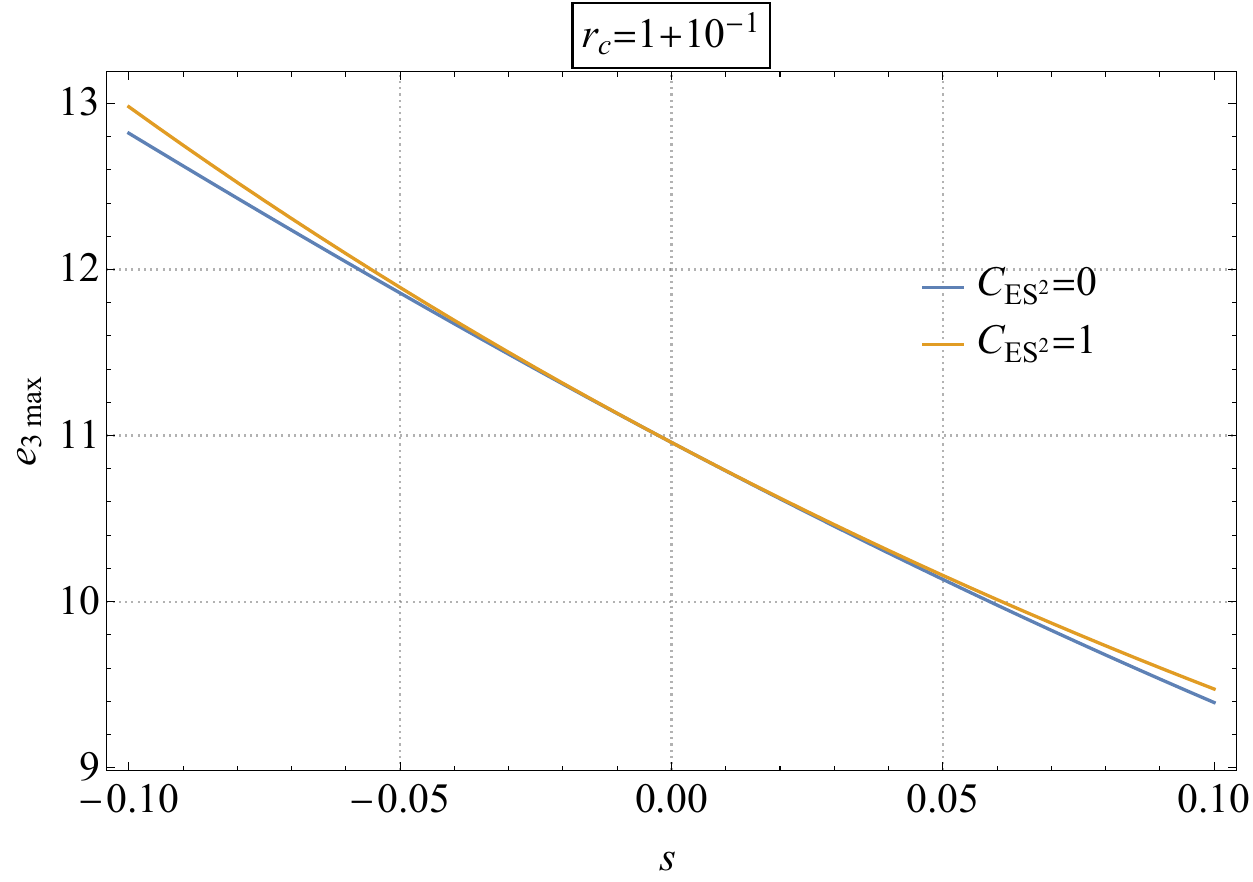}
 \includegraphics[width=0.48\textwidth]{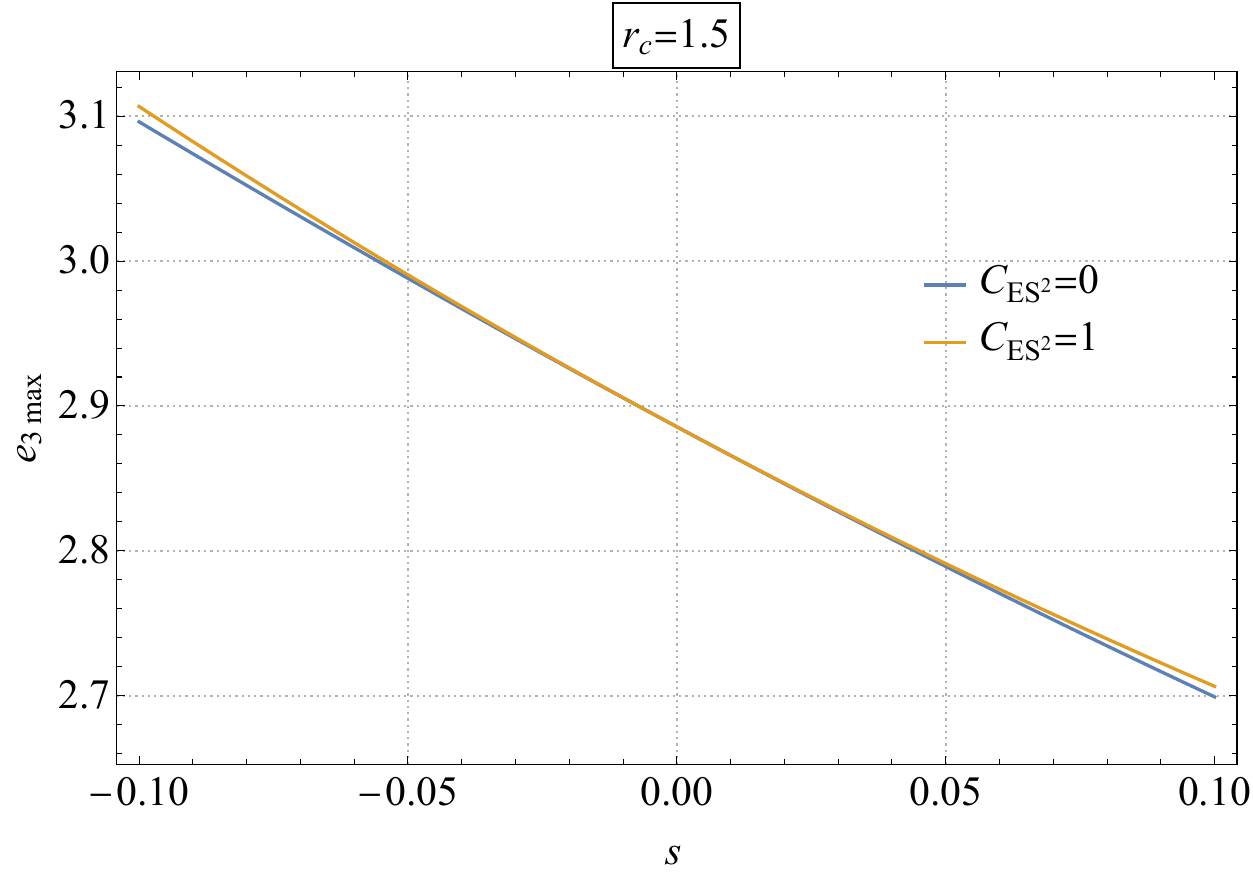}
    \caption{Maximum energy extracted by the collisional Penrose process as a function of the spin parameters $s$ at four fixed collisional points.}
    \label{fig3}
\end{figure*}

\section{numerical result}\label{sec4}
In this section, we { are going to calculate the maximum energy that can be attained via the collisional Penrose process} for different spin $s$ and the quadrupolar parameter $C_{ES^2}$. {As stated above, we fix the black hole mass at $M=1$ and the black hole angular momentum at $a=1$.} It is worth noting that the ergosphere will be located at $r_s=2$.

First, in Fig. \ref{fig1}, we present the maximal extractable energy as a function of $r_c$ for $s=0.01,\,C_{ES^2}=1$ when $j=1$ and $j=2.22$. The maximum extractable energy is unbounded near the horizon when $j=1$, i.e. the super Penrose process occurs, while the collision cannot happen near the horizon when $j=2.22$ and the maximum extractable energy is bounded. This result is similar to the previous one which considered the point particles \cite{Berti:2014lva} and particles with spin \cite{Liu:2018myg,Zhang:2018gpn}. This implies that the super Penrose process will happen for $j=1$ when we consider the quadrupolar effect with parameter $C_{ES^2}=1$. Therefore, we would like to focus on the case where $j=1$ in the following discussion and aim to find how the quadrupolar momentum affects the maximum extractable energy, which is equivalent to the maximum energy extract efficiency.

Next, in Fig.\ref{fig2}, we show that the maximum extractable energy varies with $r_c$ for $4$ different spin parameters, including $s=0.05$ for the left top panel, $s=0.1$ for the right top panel, $s=-0.05$ for the left bottom panel and $s=-0.1$ for the right bottom panel. In each panel, we fix $s$ and draw two curves for $C_{ES^2}=0$ and $C_{ES^2}=1$, respectively. It can be found that the curve representing $C_{ES^2}=1$ lies above $C_{ES^2}=0$ in every subfigure, which means the maximum extractable energy increase when the quadrupolar effect is taken into account for this $ s$. We can also find that the difference between $C_{ES^2}=0$ and $C_{ES^2}=1$ is more obvious when the absolute value of $s$ is larger. 

Then, to make the above conclusion more convincing and suitable for every spin parameter $s$, in each subfigure of Fig. \ref{fig3}, we fix the collisional point $r_c=1+\delta$ and give the maximum extractable energy varying with $s$ for $C_{ES^2}=0,\,1$. Comparing these four subfigures, we can find that, as the collisional point is away from the horizon, the maximal energy extraction will decrease and is approximately proportional to $1/\delta$. If we concentrate on a single subfigure in Fig. \ref{fig3}, we can find that the maximum extractable energy varies continuously with $s$ and monotonically decreases with $s$.  Further, the $C_{ES^2}=1$ curve stays above no matter what $s$ is. And the curve for $C_{ES}=0$ and for $C_{ES^2}=1$ only tangent to each other when $s=0$. This agrees with the result getting in Fig.\ref{fig2}.

These findings highlight the necessity of considering the quadrupolar approximation and the influence of particle shape in studying the collisional Penrose process.
{ The value of $C_{ES^2}$ represents the quadrupolar deformations induced by spin. A larger $C_{ES^2}$ implies bigger deviation from point particle.}
This suggests that the Penrose process that occurs in reality may be enhanced and there might be higher energy extraction efficiency from the rotating black hole.

\section{conclusion}\label{sec5}

We investigated the collisional Penrose process in the vicinity of an extremal Kerr black hole using the pole-dipole-quadrupole approximation to the extended tested particles. We considered the motion of test particles within the ergosphere {of the black hole} and analyzed the energy extraction of the process near the horizon.  Taking the quadrupolar approximation into account allows for a more realistic understanding of energy extraction near rotating black holes. {We found} that the super Penrose process happens with the quadrupolar effect and the maximum energy extracted in the collisional Penrose process is influenced by both the spin parameter and quadrupole parameter {of the particle}. When the quadrupolar parameter $C_{ES^2}$ is fixed, the energy extracted efficiency will decrease as the spin parameter $s$ increase. When the spin parameter $s$ is fix, the energy extraction efficiency for $C_{ES^2}=1$ case will be higher than that for $C_{ES^2}=0$ case.

Although we have determined that the maximum energy extraction will increase and the super Penrose process will occur when we consider the quadrupolar effect, there are still many unresolved questions.

Firstly, $C_{ES^2}$ is one of the quadrupolar parameters that describes the quadrupolar deformations induced by spin. Besides $C_{ES^2}$, there are other parameters to describe the quadrupolar deformations of the particles, such as the quadrupolar deformations induced by tidal force \cite{Steinhoff:2012rw}. It is also interesting to study the energy extraction influenced by other parameters.
Then, we used the partial's equation of motion given by pole-dipole-quadrupole approximation, which neglects the back reaction of the test particle to the spacetime geometry. These back reactions should be considered in a more rigorous investigation.

\begin{acknowledgments}
MZ is supported by the National Natural Science Foundation of China with Grant No. 12005080. JJ is supported by the National Natural
Science Foundation of China with Grant No. 210510101, the Guangdong Basic and Applied Research Foundation with Grant No. 217200003, and the Talents Introduction Foundation of Beijing Normal University with Grant No. 310432102.
\end{acknowledgments}

\bibliography{penrosequadrupole}

\end{document}